\documentstyle[floats,prd,aps,epsf]{revtex}

\draft
\begin{document}

\def\agt{\mathrel{\raise.3ex\hbox{$>$}\mkern-14mu\lower0.6ex\hbox{$\sim$}}}
\def\alt{\mathrel{\raise.3ex\hbox{$<$}\mkern-14mu\lower0.6ex\hbox{$\sim$}}}

\newcommand{\beq}{\begin{equation}}
\newcommand{\eeq}{\end{equation}}
\newcommand{\beqn}{\begin{eqnarray}}
\newcommand{\eeqn}{\end{eqnarray}}
\newcommand{\pa}{\partial}
\newcommand{\vp}{\varphi}
\newcommand{\varep}{\varepsilon}
\newcommand{\ep}{\epsilon}
\newcommand{\comp}{(M/R)_\infty} 
\def\bI{\hbox{$\,I\!\!\!$--}}

\twocolumn[\hsize\textwidth\columnwidth\hsize\csname
@twocolumnfalse\endcsname


\begin{center}
{\large\bf{Robustness of a high-resolution central scheme for hydrodynamic
simulations \\
in full general relativity 
}}
~~\\
~~\\
Masaru Shibata$^1$ and Jos\'e A. Font$^2$ 
~~\\
~~\\
{\em $^1$ Graduate School of Arts and Sciences, 
University of Tokyo, Komaba, Meguro, Tokyo 153-8902, Japan \\
$^2$ Departamento de Astronom\'{\i}a y Astrof\'{\i}sica,
Universidad de Valencia, 
46100 Burjassot (Valencia), Spain
}
\end{center}

%
%
%
\begin{abstract}
A recent paper by Lucas-Serrano et al.~\cite{lucas} indicates that a
high-resolution central (HRC) scheme is robust enough to yield
accurate hydrodynamical simulations of special relativistic flows in
the presence of ultrarelativistic speeds and strong shock waves. In
this paper we apply this scheme in full general relativity (involving
{\it dynamical} spacetimes), and assess its suitability by performing
test simulations for oscillations of rapidly rotating neutron
stars and merger of binary neutron stars. It is demonstrated that this 
HRC scheme can yield results as accurate as 
those by the so-called high-resolution shock-capturing (HRSC) 
schemes based upon Riemann solvers. Furthermore, the adopted HRC scheme 
has increased computational efficiency as it avoids the costly solution 
of Riemann problems and has practical advantages in the modeling of 
neutron star spacetimes. Namely, it allows simulations with stiff 
equations of state by successfully dealing with very low-density unphysical
atmospheres. These facts not only suggest that such a HRC scheme may
be a desirable tool for hydrodynamical simulations in general relativity, 
but also open the possibility to perform accurate magnetohydrodynamical 
simulations in curved dynamic spacetimes. 
\end{abstract}


\vskip2pc]

Hydrodynamics simulation in general relativity (GR) is the best
theoretical approach for investigating dynamical phenomena in
relativistic astrophysics such as stellar core collapse
to a neutron star and a black hole, and the merger of binary neutron
stars. In the past several years this field has witnessed major
development, to the stage that it is now feasible to perform accurate
simulations for such general relativistic phenomena (see
e.g.~\cite{shiba99,font02,ref1,STU,STU2}). Currently, the most favored
approach to hydrodynamics simulations in full GR 
combines the use of the so-called BSSN formalism to solve Einstein's
field equations~\cite{BSSN} and upwind high-resolution shock-capturing
(HRSC) schemes to solve the hydrodynamics equations~\cite{Toni} in
conservation form. Hereafter, HRSC schemes are referred to as those in
which the hydrodynamics equations are solved by means of (either exact
or approximate) Riemann solvers~\cite{MM,Toni} (i.e.~Godunov-type
schemes).

Regarding the solution of the hydrodynamics equations it has been
shown in a few recent papers~\cite{ref2,lucas} that high-resolution
central symmetric schemes (HRC scheme hereafter) yield numerical
solutions as accurate as those by HRSC schemes for special
relativistic flows (see e.g.~\cite{toro} for a general introduction to
HRSC and HRC schemes). The main conclusion of those works highlights
the importance of the {\em conservation form} of the adopted scheme
(either upwind or central) in conjunction with high-order
cell-reconstruction procedures (to compute the numerical
hydrodynamical fluxes at cell interfaces) to gain accuracy while
reducing as much as possible the inherent diffusion of central schemes
at discontinuities. It is well-known that if a numerical scheme
written in conservation form converges, it automatically guarantees
the correct Rankine-Hugoniot (jump) conditions across discontinuities. 
This shock-capturing property is hence shared by both upwind and
symmetric schemes. For practical reasons the most appealing feature of
HRC schemes is the fact that, contrary to upwind HRSC schemes, they
entirely sidestep the use of Riemann solvers, which results in a great
simplification for their numerical implementation as well as in
enhanced computational efficiency. However, it has not yet been
clarified whether HRC schemes can also yield numerical results as
accurate as those of HRSC schemes for simulations in full
GR involving {\it dynamical} spacetimes.

The aim of this paper is to demonstrate the robustness of a particular
HRC scheme proposed by~\cite{kurganov-tadmor}, and first used in
special relativistic hydrodynamics by~\cite{lucas}, for 
problems in full GR. As we have done in previous papers
(e.g.~\cite{shiba99,font02,shiba2d}), test simulations in both
axisymmetry (rotating neutron stars) and full three-dimension
(binary neutron star mergers) are performed to assess this fact. 

The numerical simulations are carried out using the same mathematical
formulation as in~\cite{STU}, to which the interested reader is
addressed for details about the basic equations, the gauge conditions,
and the computational method. Einstein's evolution equations are
solved using the so-called BSSN formalism \cite{BSSN}, adopting a
slight variation of the original form of the equations, which is
reported in \cite{STU}. The hydrodynamics equations are written in
conservation form and solved using both a Roe-type HRSC
scheme~\cite{shiba2d} and a HRC scheme~\cite{lucas}, with either
the PPM third-order cell-reconstruction or the MC slope
limiter. Violations of the Hamiltonian constraint and conservation
of ADM mass and angular momentum are monitored to check the accuracy
of the simulations. 

We use a fixed uniform grid for both the axisymmetric and the
three-dimensional (3D) simulations. The former are carried out in
cylindrical coordinates ($\varpi, z$) assuming equatorial plane
symmetry. Computational grids of size $(N+1, N+1)$ with $N=90$, 
120, 180, 240, and 360 are used, with which convergence is shown. 
The 3D simulations are performed in Cartesian coordinates 
assuming equatorial plane symmetry as well. In this case the grid 
adopted in the present test simulations consists of (377,377,189) 
zones for $(x, y, z)$ respectively. 

In the axisymmetric simulations of isolated rotating neutron stars a
$\Gamma$-law equation of state (EOS) is used, i.e.~$P=(\Gamma-1)\rho
\varep$. Here, $P$ is the pressure, $\rho$ the rest-mass density,
$\varep$ the specific internal energy, and $\Gamma$ the adiabatic
constant for which we choose the values 2 and 2.5. The initial
conditions for the equilibrium models are built using a polytropic EOS
$P = K \rho^{\Gamma}$, where $K$ is the polytropic constant.

Correspondingly, for the 3D simulations of binary
neutron star merger a hybrid EOS is adopted, as described
in~\cite{STU2}. In this EOS, the pressure and the specific internal
energy are written in the form $P=P_{\rm cold}(\rho) + P_{\rm th}$ and
$\varep=\varep_{\rm cold}(\rho) + \varep_{\rm th}$ where $P_{\rm
cold}$ and $\varep_{\rm cold}$ are the cold (zero-temperature) parts,
and are functions of $\rho$ only. On the other hand $P_{\rm th}$ and
$\varep_{\rm th}$ are the thermal (finite-temperature) parts. During
the simulation, $\rho$ and $\varep$ are evolved, and thus $\varep_{\rm
th}$ is determined by $\varep-\varep_{\rm cold}$. For $P_{\rm th}$, we
simply set $P_{\rm th}=(\Gamma_{\rm th}-1)\rho \varep_{\rm th}$ with
$\Gamma_{\rm th}=2$. For the cold part of the hybrid EOS we use
realistic EOS for zero-temperature nuclear matter, more precisely the
SLy EOS~\cite{HP}.

As customary in grid-based hydrodynamics codes an artificial low-density
atmosphere needs to be used in those regions outside the star
representing vacuum. The density has to be low enough so that its
presence does not affect the actual dynamics of the star. In previous
simulations using a Roe-type HRSC scheme, a uniform density atmosphere
as low as $\rho_{\rm atm}=10^{-6} \rho_{\rm max}$ was used, where
$\rho_{\rm max}$ is the maximum density. (For soft EOS this value can
be much smaller; e.g., $\rho_{\rm atm} \leq 10^{-12} \rho_{\rm max}$
for $\Gamma=4/3$.) Lower values for the density could result in
numerical instabilities developing around the stellar surface.
However, we have found that when using the HRC scheme the threshold
density in the atmosphere can be {\it much smaller}. The results
presented next for the HRC scheme correspond to $\rho_{\rm atm}=10^{-10} 
\rho_{\rm max}$, irrespective of the EOS used.

\begin{figure}[thb]
\vspace{-4mm}
\begin{center}
\epsfxsize=2.8in
\leavevmode
\epsffile{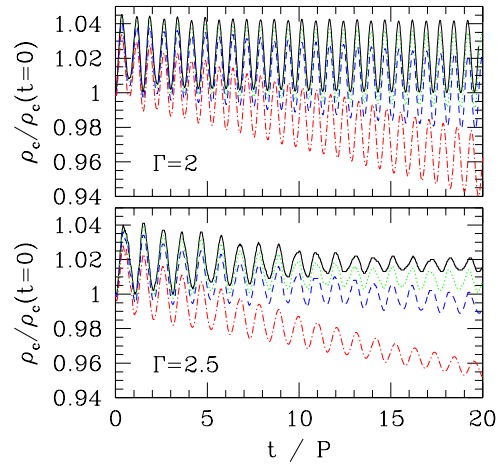} 
\end{center}
\vspace{-6mm}
\caption{ Evolution of the central density in units of the initial
value for the two rotating neutron star models considered. Time is
shown in units of the rotational period of the neutron stars $P$. The
dotted-dashed, dashed, dotted, and solid curves denote the results
with (121,121),
(181,181), (241,241), and (361,361) grid resolutions, respectively.
\label{FIG1} }
\end{figure}

\begin{figure}[thb]
\vspace{-4mm}
\begin{center}
\epsfxsize=2.8in
\leavevmode
\epsffile{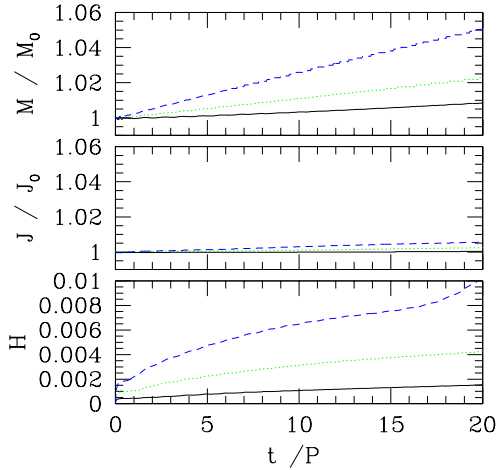}
\end{center}
\vspace{-6mm}
\caption{Evolution of the ADM mass (top), angular momentum (middle),
and violation of the Hamiltonian constraint (bottom)
for $\Gamma=2.5$. Time is shown in units of the
rotational period of the neutron stars. The dashed, dotted, and solid
curves denote the results with (91,91), (121,121), and (181,181)
grid resolutions, respectively.
\label{FIG2} }
\end{figure}

We start discussing axisymmetric simulations of oscillations of
rotating neutron stars. For these simulations we build rapidly
rotating neutron stars with uniform angular velocity. This velocity is
chosen so that it reaches the Kepler (mass-shedding) limit at the
equatorial stellar surface.

Two rotating neutron star models are considered. In one case
$\Gamma=2$ and the baryon rest mass $M_*$ is 90\% of the maximum
allowed value for uniformly rotating neutron stars of identical
EOS. This model is the same as model R2 in Ref.~\cite{shiba2d}, which
allows for a direct comparison. The other model corresponds to
$\Gamma=2.5$ and $M_*$ is 95\% of the maximum allowed mass. This is a
very compact model, since the compactness parameter, defined as
$GM/Rc^2$ where $M$ and $R$ are the ADM mass and circumferential
radius around the equatorial surface, is 
0.214. For both models, the axis ratio of polar radius to equatorial
radius is about 0.6. The ratio of the coordinate radius of
the outer boundary of the computational grid to the stellar coordinate
radius at the equator is 3. The simulations are started by reducing
the pressure by 1\% uniformly.

Figure~\ref{FIG1} shows the time-evolution of the central density for
these two models obtained using the HRC scheme for the hydrodynamics
equations. Each curve corresponds to a different grid resolution as 
explained in the caption. It is found that the HRC scheme succeeds in
keeping the stars in equilibrium in such a dynamical spacetime. The
neutron star oscillations can be followed accurately for more than 20
rotation periods. With small grid sizes (dotted and dashed lines), 
the density experiences a secular drift, decreasing with time gradually. 
The reason is that the angular momentum of the star is transported 
outward by numerical diffusion. However, this drift decreases with 
improved grid resolution, and with the highest resolution the average 
value of the central density is kept approximately constant. 
Second-order convergence is also achieved. 

It is worth to emphasize that despite the use of an artificial atmosphere 
of tiny density, the HRC scheme makes it possible to follow the evolution 
of compact neutron stars with stiff EOS with $\Gamma=2.5$ and to compute 
their fundamental oscillation frequency. Such simulation has not yet been 
accurately performed with HRSC schemes. 

In Fig.~2, we show the evolution of the ADM mass, angular momentum,
and the averaged violation of the Hamiltonian constraint (in which the
baryon rest mass density is used for the weight; see \cite{STU} for
definition) for $\Gamma=2.5$. (Similar results are obtained for 
$\Gamma=2$.) The figure shows that the conserved quantities remain
conserved to high accuracy, particularly for the finest grid, and that 
the violation of the Hamiltonian constraint remains small. The
outstanding feature is that the departure from angular momentum
conservation with the HRC scheme is much smaller than with the HRSC 
scheme (see e.g.~Fig.~5 in~\cite{shiba2d}). In the previous implementation 
the angular momentum gradually increases with time mainly due to the 
numerical error generated around the stellar surface and in the low-density 
atmosphere for which an artificial friction term was added to stabilize 
the computation. With the HRC scheme, such a drift in the angular momentum 
conservation is suppressed within 0.1\% error after 20 rotation periods 
for grid resolutions with $N \geq 180$, probably due to less numerical 
inaccuracies around the stellar surface. These results indicate that 
the HRC scheme used is a robust scheme for the simulation of isolated 
neutron stars. 

\begin{figure}[thb]
\vspace{-4mm}
\begin{center}
\epsfxsize=2.8in
\leavevmode
\epsffile{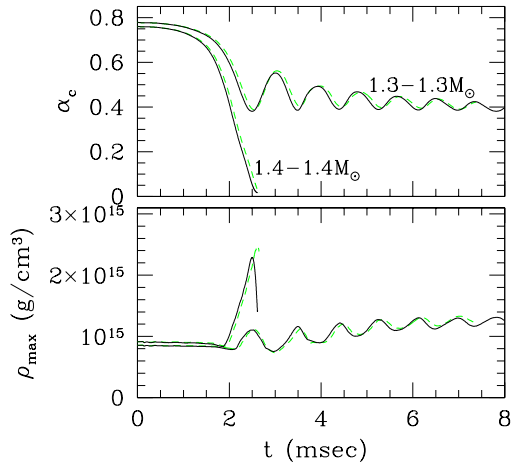}
\end{center}
\vspace{-6mm}
\caption{Evolution of the maximum value of $\rho$ and the central
value of $\alpha$ for the merger of equal-mass binary neutron stars,
1.3--1.3$M_{\odot}$ and 1.4--1.4$M_{\odot}$. In the former and latter
cases, a massive neutron star and a black hole are the end-products of
the merger, respectively. 
\label{FIG3} }
\end{figure}

\begin{figure}[thb]
\vspace{-4mm}
\begin{center}
\epsfxsize=2.8in
\leavevmode
\epsffile{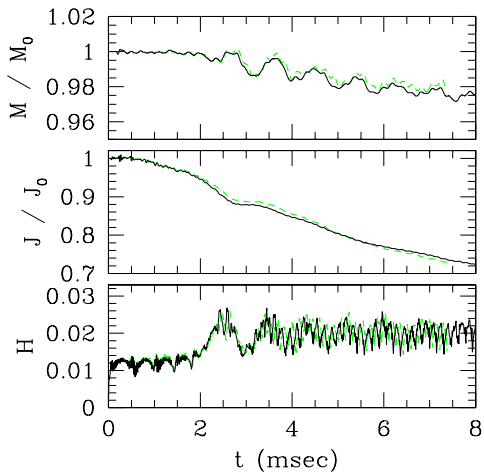}
\end{center}
\vspace{-6mm}
\caption{Evolution of the ADM mass (top), angular
momentum (middle), and violation of the Hamiltonian constraint
(bottom) for the merger of binary neutron stars with masses
1.3--1.3$M_{\odot}$. The solid and dashed curves in all panels
indicate the results obtained with the HRC scheme and with the HRSC
scheme, respectively.
\label{FIG4} }
\end{figure}

We now turn to present the results 
of numerical simulations of binary neutron star mergers. In the
present test we choose binaries of equal mass with 1.3--1.3$M_{\odot}$
and 1.4--1.4$M_{\odot}$. As found in \cite{STU2} using a HRSC scheme,
a massive neutron star and a black hole are formed for the former and
latter cases, respectively.

In Fig.~3, we show the time-evolution of the central value of the
lapse function, $\alpha_c$, and the maximum value of the density,
$\rho_{\rm max}$, for these two models. The solid and dashed curves
indicate the results obtained with the HRC and HRSC schemes,
respectively. In the smaller mass case, a massive neutron star is
formed after the merger, and hence, $\alpha_c$ and $\rho_{\rm max}$
show a series of small-amplitude oscillations until they eventually relax
to quasi-stationary values. For both hydrodynamical schemes the
amplitude and the frequency of the resulting neutron star oscillations
agree well with each other. On the other hand, the outcome of the
merger of the 1.4--1.4$M_{\odot}$ binary is a black hole, as can be
directly inferred from the rapid collapse of the central lapse and the
rapid growth of the maximum density (from $t\sim 2$ ms onwards). Black
hole formation is signaled by the appearance of an apparent horizon,
which is detected in both implementations. In particular the time of
formation of the apparent horizon agrees approximately for both
schemes, with a time difference of about 0.07 ms. For the two binary
mergers considered, a small time lag in the evolution of $\alpha_c$ and
$\rho_{\rm max}$ is observed between the two results computed by the
different schemes. Its origin is likely the difference in the magnitude 
of the friction term around the stellar surface already discussed 
before which could generate an error in the angular momentum 
conservation. As mentioned above, this error is smaller with the HRC 
scheme. 

In Fig.~4, we show the evolution of the ADM mass, angular
momentum, and averaged violation of the Hamiltonian constraint for the
1.3--$1.3M_{\odot}$ binary merger. In this case, the ADM mass and
angular momentum of the system show a gradual decrease due to the
emission of gravitational waves~\cite{STU2}. Again, it is found that 
the two results agree well with each other within $\sim 0.5\%$.
The averaged violation of the Hamiltonian constraint remains
approximately of identical magnitude, $\sim 0.02$, which indicates
that the accuracy of the results of the two hydrodynamical schemes is
approximately identical.

To summarize, it has been shown through simulations of pulsating and 
rotating neutron stars, and binary neutron star mergers, that the 
results produced by the HRC scheme proposed by~\cite{kurganov-tadmor} 
agree well with those obtained with a Roe-type HRSC scheme. The accuracy 
measured by the evolution of the ADM mass, angular momentum, and 
violation of the Hamiltonian constraint in the HRC scheme are as good 
as or even better than those obtained for the HRSC scheme. In addition, 
the HRC scheme has a number of advantages to the HRSC scheme: (1) it 
is straightforward to implement since the solution of Riemann 
problems is avoided; hence one does not need to compute the complicated 
sets of eigenvectors of the Jacobian matrices associated with the fluxes
(transport terms) of the hydrodynamics equations; (2) for this 
reason the computational costs of the HRC scheme are much less
expensive, as the characteristic information required in HRSC schemes
is not necessary. In the tests reported in this paper we have found
that in our fully general relativisitc implementation, the
computational time is saved by about 20 \%; (3) the density of the
unphysical atmosphere one needs to build around isolated stars when
adopting the conservative form of the hydrodynamics equations can be
several orders of magnitude smaller than that in HRSC
schemes. Associated with this advantage the code can be applied for
neutron stars with a large adiabatic index $\Gamma =2.5$.

These facts illustrate that HRC schemes can be useful and robust tools
for hydrodynamical simulations in full GR involving {\it
dynamical} spacetimes. In addition, their suitability over HRSC
schemes becomes further apparent when the wave structure of the
hyperbolic system to solve is unknown, as it is partially the case in
general relativistic magnetohydrodynamics (GRMHD).  Hence, HRC schemes
can help the achievement of GRMHD simulations in which the equations
to solve are more complicated than those of purely hydrodynamical
flows \cite{GRMHD}.

{\em Acknowledgments}: 
The numerical simulations were performed on the FACOM VPP5000 computers 
at the data processing center of NAOJ. This work was in part supported by 
Monbukagakusho Grant (Nos. 15037204, 15740142, and 16029202) and by the 
Spanish Ministerio de Ciencia y Tecnolog\'{\i}a (AYA2004-08067-C03-C01).


\begin{thebibliography}{99}

\bibitem{lucas} 
A. Lucas-Serrano, J. A. Font, J. M. Ib\'anez, and J. M. Mart\'{\i}, 
Astron. Astrophys. {\bf 428}, 703 (2004). 

\bibitem{shiba99}
M. Shibata, Phys. Rev. D {\bf 60}, 104052 (1999). 

\bibitem{font02} J. A. Font et al., Phys. Rev. D {\bf 65}, 084024 (2002). 

\bibitem{ref1} 
M. Shibata and K. Ury\=u, Phys. Rev. D {\bf 61}, 064001 (2000); 
Prog. Theor. Phys. {\bf 107}, 265 (2002): 
M. Miller, P. Gressman, and W.-M. Suen, Phys. Rev. D {\bf 69}, 064026 (2004): 
M. D. Duez, P. Marronetti, T. W. Baumgarte, and 
S. L. Shapiro, Phys. Rev. D {\bf 67}, 024004 (2003): 
L. Baiotti et al., Phys. Rev. D {\bf 71}, 024035 (2005). 

\bibitem{STU} M. Shibata, K. Taniguchi, and K. Ury\=u,
Phys. Rev. D {\bf 68}, 084020 (2003).

\bibitem{STU2} M. Shibata, K. Taniguchi, and K. Ury\=u,
Phys. Rev. D {\bf 71}, 084013 (2005). 

\bibitem{BSSN} M. Shibata and T. Nakamura, Phys. Rev. D {\bf 52}, 5428 (1995):
T. W. Baumgarte and S. L. Shapiro, Phys. Rev. D {\bf 59}, 024007 (1999). 

\bibitem{Toni} J. A. Font, Living Rev. Relativity {\bf 6}, 4 (2003). 

\bibitem{MM} J. M. Mart\'{\i} and E. M\"uller, Living Rev. Relativity {\bf 6},
7 (2003). 

\bibitem{ref2} L. Del Zanna and N. Bucciantini, Astron. Astrophys. 
{\bf 390}, 1177 (2002): P. Anninos and P. C. Fragile, Astrophys. J. Suppl.
{\bf 144}, 243 (2003). 

\bibitem{toro} E.~F. Toro, {\em 
Riemann Solvers and Numerical Methods for Fluid Dynamics
} (Springer Verlag, 1997).

\bibitem{kurganov-tadmor} 
A. Kurganov and E. Tadmor, J.~Comput.\ Phys. {\bf 160}, 214 (2000).

\bibitem{shiba2d} M. Shibata, Phys. Rev. D {\bf 67}, 024033 (2003).



\bibitem{HP} P. Haensel and A. Y. Potekhin,
Astron. Astrophys. to be published (astro-ph/0408324), and
references therein. 

\bibitem{GRMHD} 
M. D. Duez, Y.-T. Liu, S. L. Shapiro, and B. Stephens,
submitted to Phys. Rev. D; L.~Ant\'on et al, in preparation;
M. Shibata and Y. Sekiguchi, in preparation. 

\end{thebibliography}
\end{document}